# Signal amplification in simple metal-insulator transition devices

*Victor Palin[1], Nareg Ghazikhanian[1], Matthew Frame[2], Yayoi Takamura[2], Ivan K. Schuller[1], Pavel Salev[3]*

[1]*Department of Physics, University of California San Diego*
[2]*Department of Materials Science & Engineering, University of California Davis*
[3]*Department of Physics & Astronomy, University of Denver*

**Signal dissipation in large-scale neural networks can lead to information loss and ultimately to computational failures, necessitating local signal amplification at neuron – synapse connections [1]. In biological nervous systems, axons are responsible for local signal amplification [2,3]. Translating axon functionality into hardware, i.e., the ability to amplify and transmit signals without loss, is non-trivial because emulating the human brain implies building networks composed of ~10 billion interconnected neurons, each requiring a dedicated compact and scalable amplifier [4]. Here, we demonstrate signal amplification in simple two-terminal devices made of a metal-insulator transition material. By operating the devices on the verge of the phase transition and taking advantage of negative differential resistance, we achieve robust signal amplification up to a factor of ~11.5. We also demonstrate the amplification of spiking sequences generated by a real neuristor, opening new exciting opportunities for the direct integration of artificial neurons and axons. The amplification can be controlled by easily adjustable experimental parameters, including DC bias, AC excitation, series resistance, and temperature. We further propose a model that predicts the gain using readily observable transport characteristics. Our results establish a framework for developing and optimizing axon-like amplification functionalities in nonlinear electronic materials.**

## Introduction

The human brain is a highly interconnected network composed of ~$10^{10}$ neurons and ~$10^{14}$ synapses, enabling fast, highly parallel, and energy-efficient information processing [5,6]. The reliable operation of such a large network critically depends on the transmission of signals between neurons with minimal loss. A key element for achieving high signal transmission efficiency is the axon, whose primary functionality is to regenerate signals that propagate in a lossy medium. This regeneration is achieved via the coordinated activity of voltage-gated ionic channels located at the nodes of Ranvier, which control $Na^+$ and $K^+$ concentrations along the axon path, allowing action potentials to travel over long distances without diminishing in strength [7]. Electrochemical signal amplification in axons ensures robust information transmission across large-scale biological neural networks.

Extensive theoretical and experimental efforts have been conducted to identify materials with properties suitable for emulating brain-like functions, especially in correlated oxides [8–12]. Hardware implementation of neural networks requires finding basic elements that mimic the functionalities of neurons, synapses, and axons [13–20]. Volatile resistive switching in phase-change materials and select ionic migration systems, as well as resistance/magnetization oscillations in spin Hall nano-oscillators, enable simple devices that replicate various neuronal behaviors [8,21–27]. Multilevel nonvolatile resistive switching in electrochemical switches, oxide memristors, 2D materials, ferroelectric tunnel junctions, and spintronic memories provides opportunities to emulate synaptic plasticity [8,26–31]. The above neuronal and synaptic implementations rely on tunable resistive elements that are inherently lossy, which inevitably results in a fast decay of information in large-scale networks. Consequently, signal attenuation is a critical limitation for large-scale hardware neural networks, highlighting the urgent need to implement axon-like functionalities to locally amplify signals at neuron-synapse junctions. Complementing each neuron device with a conventional CMOS amplifier is unlikely to yield a feasible solution because of the large number of neurons, e.g., ~$10^{10}$ neurons in an artificial network that has a complexity comparable to the human brain [32]. A new approach is needed to design compact, scalable, and energy efficient axon-like amplifiers that are also compatible with other neuromorphic materials.

In contrast to neuronal and synaptic functionalities, signal amplification, i.e., axon-like behavior, remains largely unexplored. It has been shown that a modest amplification of ~1.6× can be achieved in a spin crossover material, $LaCoO_3$, that is electrically biased at the edge of chaos (EOC), where small fluctuations are amplified within a certain bound without growing into chaotic dynamics [33]. The EOC is an exotic state whose stabilization requires a unique combination of material properties and careful circuit design, making practical device optimization challenging [34].

Here, we demonstrate amplification using a non-chaotic metal-insulator transition (MIT) material. MIT materials are well-known for their ability to reproduce neuronal spiking [22,35–39], providing a natural platform for integrating our axon-like MIT amplifier with artificial neuristors. We achieved a large gain of ~11.5× by electrically triggering the MIT and utilizing stable and deterministic insulator-phase injection into an otherwise uniform metallic-phase matrix. The gain of the MIT amplifier can be controlled by several easily adjustable parameters, such as DC bias, AC voltage, series resistance, and temperature. The MIT amplifier is implemented in a scalable two-terminal device that can achieve stable amplification without the need for a precisely defined transmission line. Finally, we propose a simple model that accurately predicts the MIT amplifier gain using the device's I–V characteristic as an input, which can help streamline practical device optimization by relying on easily observable material parameters.

## Results

To investigate amplification in MIT materials, we patterned planar two-terminal devices on a 20-nm-thick $La_{0.7}Sr_{0.3}MnO_3$ (LSMO) film epitaxially grown on a (001)-oriented $SrTiO_3$ substrate (inset of **Fig. 1a**). **Fig. 1a** shows the non-linear I-V characteristics of a 2×1 μm² device. Applying voltage induces volatile low-to-high resistance switching, which is mediated by the electrical triggering of the MIT (**Suppl. Fig. S1**) and accompanied by a pronounced N-type negative differential resistance (NDR) region (a portion of the I-V curve where the *dI/dV* slope is negative) [40,41]. We performed I-V measurements using several resistors ($R_s$) connected in series with the LSMO device. Adding the series resistance allows tuning of the I-V nonlinearity and makes the volatile resistive switching more abrupt and even hysteretic (indicated by up/down arrows in **Fig. 1a**). As we show later in the paper, tuning I-V nonlinearities is one of the key parameters for achieving high amplification.

We used a simple voltage-divider-type circuit, consisting of a conventional resistor $R_s$ and an LSMO device, to amplify small AC excitations (**Fig. 1b**). In this voltage divider, applied electrical signals are dynamically redistributed between the constant series resistor $R_s$ and the voltage-dependent resistance of LSMO, enabling amplification. The circuit is powered by a superimposed DC bias that drives the LSMO into the MIT switching regime and a small AC excitation that we aim to amplify. Two oscilloscope channels monitor the voltages in the circuit, $V_{FG}$ and $V_{out}$. We note that because our function generator (FG) has a built-in 50 Ω series resistor, we compute the input voltage as

$$V_{in} = V_{FG} + (50\ \Omega) \cdot I \qquad (1)$$

where $I = (V_{FG} - V_{out})/R_s$ is the current in the circuit. Accordingly, we report the effective series resistance throughout the paper as $(R_s + 50\ \Omega)$ to account for the built-in resistance of the FG.

When the DC bias is below the MIT triggering threshold, the LSMO device acts as a nearly constant resistor. In this case, the applied AC excitation is distributed between the series resistance $R_s$ and the LSMO, and the AC component in the output signal is always smaller than the input AC excitation, i.e., the AC signal is attenuated as expected for a conventional voltage divider. When the DC bias is sufficiently high to trigger the MIT, an insulating phase barrier forms in the LSMO device, splitting the initially uniform metallic phase matrix [40,41]. The small AC excitation superimposed on the DC bias dynamically modulates the size of the voltage-dependent barrier (**Fig. 1c**), resulting in a dynamic variation of the LSMO resistance. When the LSMO resistance changes, an additional AC current is generated in the electrically biased circuit. This extra AC signal adds to the input AC excitation, which is the basis for achieving amplification. Because the change of the LSMO resistance is due to the phase transition, even a small AC excitation can lead to a large response, enabling high-gain amplification.

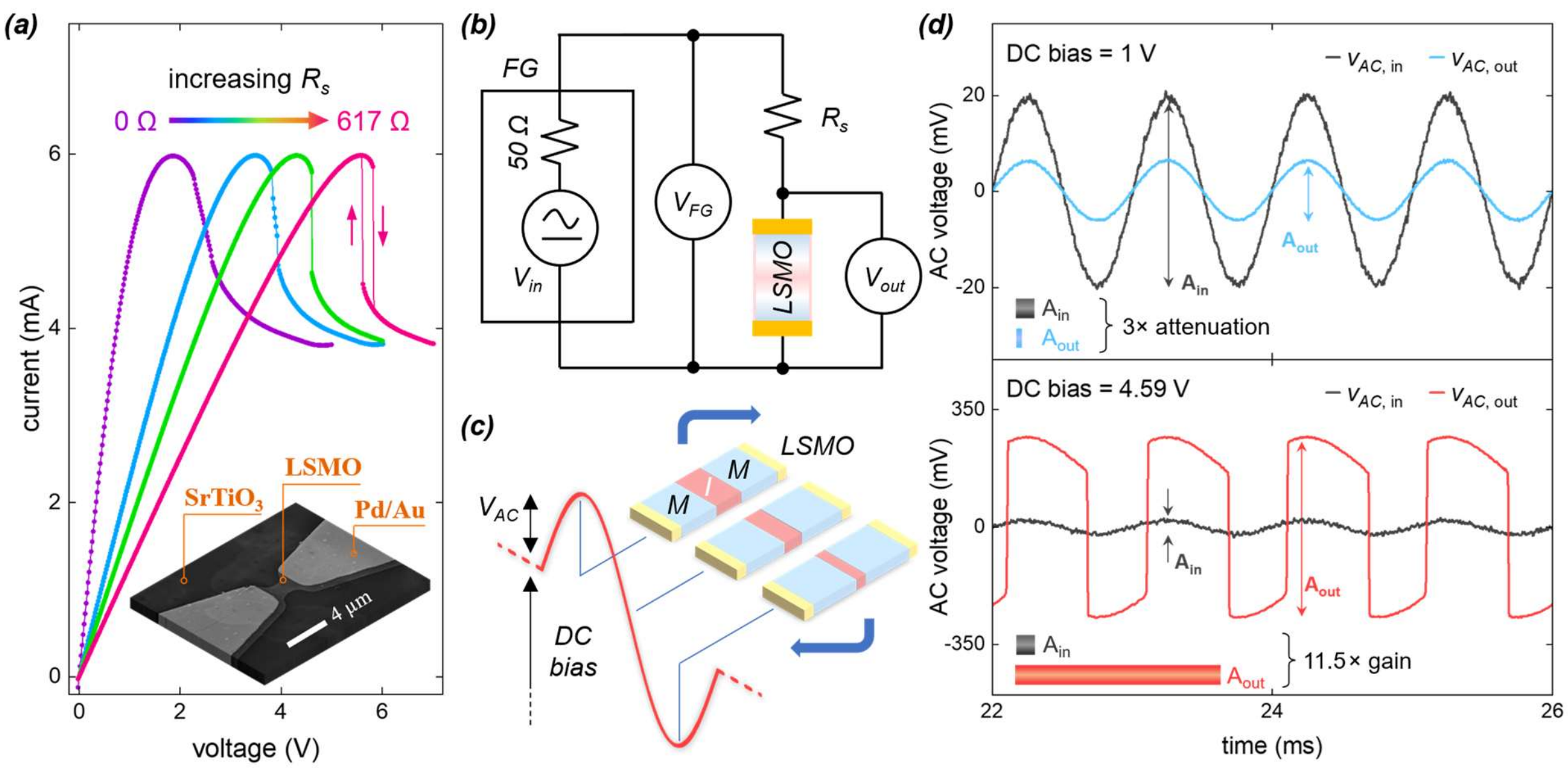

**Fig. 1. a.** I-V characteristics recorded using several series resistors, $R_s$. All curves display a characteristic NDR and a rapid MIT switching, which can be tuned by varying $R_s$. The inset shows an SEM image of a 2×1 μm² LSMO device. **b.** Electrical circuit used for the amplification measurements. A function generator supplies a superimposed small AC excitation + DC bias signal. Two oscilloscope channels, $V_{FG}$ and $V_{out}$, are used to monitor the voltages. **c.** Schematic illustrating the modulation of the LSMO insulating barrier size by a small AC excitation. **d.** Input and output AC signals recorded using a DC bias of 1 V (top panel) and 4.59 V (bottom panel). When the LSMO device is biased far away from the MIT switching ($V_{DC}$ = 1 V), the output signal is attenuated. When the LSMO device is biased at the MIT switching threshold ($V_{DC}$ = 4.59 V), the output signal is strongly amplified. The amplification measurements were performed at T = 80 K using $R_s$ = 400 Ω and $V_{AC,\,in}$ = 40 mV.

We achieved more than an order-of-magnitude amplification of AC signals in our LSMO devices. **Fig. 1d** shows an example of the amplifier response to a 40 mV 1 kHz AC input excitation superimposed on top of 1 V (top panel) and 4.59 V (bottom panel) DC bias. We note that **Fig. 1d** shows only the AC signal components (raw signals containing both AC and DC components are shown in **Suppl. Fig. S2**). At 1 V DC bias, the LSMO device is far away from the MIT switching, and the device acts as a conventional Ohmic resistor. Consequently, only a fraction of the input AC excitation (top panel, black line) is distributed to the output across the LSMO device (top panel, blue line), i.e., the circuit attenuates the input AC signal. In contrast, when the DC bias of 4.59 V brings the LSMO to the onset of MIT switching and the device can no longer be viewed as an Ohmic resistor, we observed a strong amplification of the input AC excitation (bottom panel, black line) at the device output (bottom panel, red line). By defining the AC signal gain as

$$AC\ gain = \frac{V_{AC\ out}}{V_{AC\ in}} \qquad (2)$$

this experiment yields a very large amplification of ~11.5×. Our results, thus, directly demonstrate that a simple voltage-divider-type circuit consisting of one linear and one nonlinear resistor produces high-gain amplification when the nonlinear resistor is biased into the volatile resistive switching regime.

It is interesting to note that the output AC signals appear distorted when our circuit is biased into the high amplification regime (**Fig. 1d**, bottom panel). The high amplification is associated with the MIT switching in the LSMO device. Consequently, the nonlinearity of the LSMO resistance is effectively imprinted onto the output AC signal. The abrupt jumps in $V_{AC,\,out}$ curve in **Fig. 1d** correspond to the abrupt jump at the MIT switching threshold in the I-V measurements in **Fig. 1a**. The output AC signal distortions can be minimized (albeit at the cost of reduced gain) by biasing the device deeper in the high-resistance state. In this regime, the LSMO still exhibits the non-Ohmic NDR behavior, but the strong nonlinearities associated with the abrupt MIT switching are no longer accessible to the small AC excitation (**Suppl. Fig. S3**).

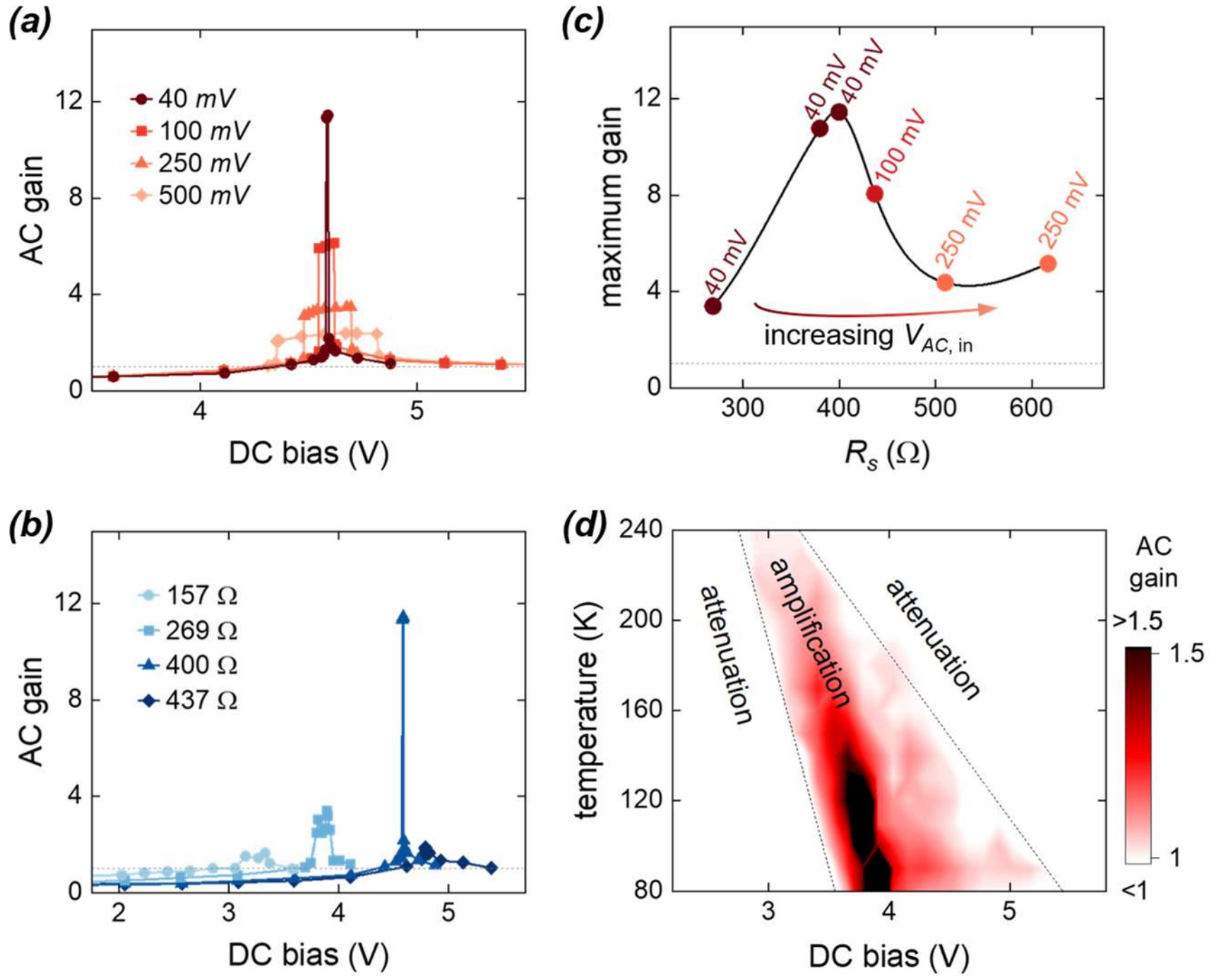

**Fig 2. a.** AC gain as a function of DC bias measured using several AC excitation amplitudes and a 400 Ω series resistance. Smaller AC excitation amplitudes maximize the gain over a narrow bias window, while larger excitations expand the DC bias window at the cost of reduced gain. **b.** AC gain as a function of DC bias measured using several series resistances and a 40-mV AC excitation. The highest gain is achieved at the optimal series resistance of 400 Ω. **c.** Maximum gain measured for various series resistance at the optimal AC excitation. Adjusting both the series resistance and the AC excitation enables high gain over a wide range of parameters. **d.** Temperature-voltage map of the AC gain. At higher temperatures, the DC bias window for amplification is narrow, but amplification can be always achieved as long as the LSMO device exhibits MIT switching and NDR. Measurements in panels (a-c) were performed at T = 80 K. Data in panel (d) were measured using $R_s$ = 400 Ω and $V_{AC,\,in}$ = 40 mV.

The amplification is controlled by several easily adjustable parameters, including DC bias, AC excitation, series resistance, and temperature. The plots in **Fig. 2** show that an AC gain greater than 1 appears within a certain DC bias range, typically at 3 – 6 V in our experiments. The DC bias window of amplification coincides with the MIT switching in LSMO and the emergence of an NDR in the I-V characteristic, indicating a close relation between the amplification and strong electrical nonlinearities in LSMO. The highest gain (up to 11.5×) is achieved at the lowest AC excitation amplitude, 40 mV in our experiments (**Fig. 2a**). This large gain, however, appears in a rather narrow DC bias window. Increasing the AC excitation amplitude reduces the gain but broadens the DC bias window over which appreciable amplifications (e.g., *AC gain* > 2) can be achieved.

The amplification has a non-monotonic dependence on the series resistance (**Fig. 2b**). The maximum gain initially increases rapidly as $R_s$ increases from 157 Ω to 400 Ω. Further increasing the series resistance, however, significantly suppresses the maximum amplification. In general, we observed that maximizing gain requires finding a balance between the sharpness of the MIT switching and the width of the switching hysteresis in the I-V characteristic (**Fig. 1a**). As demonstrated in **Fig. 2c**, this balance is not restricted to a single operation point. Simultaneously adjusting the series resistance and the AC excitation amplitude (**Suppl. Fig. S4-5**) allows for achieving relatively high-gain operation over a wide parameter range.

Amplification is achieved over a wide range of temperatures. **Fig. 2d** maps the amplification DC bias window at different temperatures. These measurements were performed using an AC excitation amplitude of 40 mV and a series resistance of 400 Ω, i.e., the conditions at which the highest AC gain was observed at 80 K. As temperature increases, the amplification DC bias window becomes narrower. Nonetheless, we found that the amplification persists as long as the LSMO device exhibits the MIT switching and NDR. Previous work showed that the MIT switching and NDR in LSMO can be found up to ~320 K [40], suggesting that further device optimization can enable the amplification operation at room temperature.

We propose a simple model that provides guidelines for selecting the MIT materials and design devices with optimal amplification characteristics. Our experiments establish a connection between the material's nonlinear electrical properties (particularly NDR and MIT switching) and amplification. **Fig. 3a** shows the experimental I-V characteristic of an LSMO device in series with a 400 Ω resistor (top panel), together with the calculated $\Delta I/\Delta V$ (bottom panel). Two regions can be identified. First, in the 0 – 4.3 V range, $\Delta I/\Delta V > 0$, indicating that the device exhibits conventional positive differential resistance (PDR). Second, in the 4.3 – 6 V range, $\Delta I/\Delta V < 0$, signifying the onset of NDR. The sharp negative $\Delta I/\Delta V$ jump corresponds to the insulating phase barrier formation in LSMO [40,41]. Comparing $\Delta I/\Delta V$ with the experimentally measured gain (**Fig. 3b**, black symbols), we notice that the maximum gain coincides with the negative peak of $\Delta I/\Delta V$. Analyzing the distribution of AC and DC signals in the circuit (see **Suppl. Mater. A** for details), we obtained an equation to estimate the gain:

$$AC\ gain = 1 - R_s \cdot \frac{\Delta I}{\Delta V} \qquad (3)$$

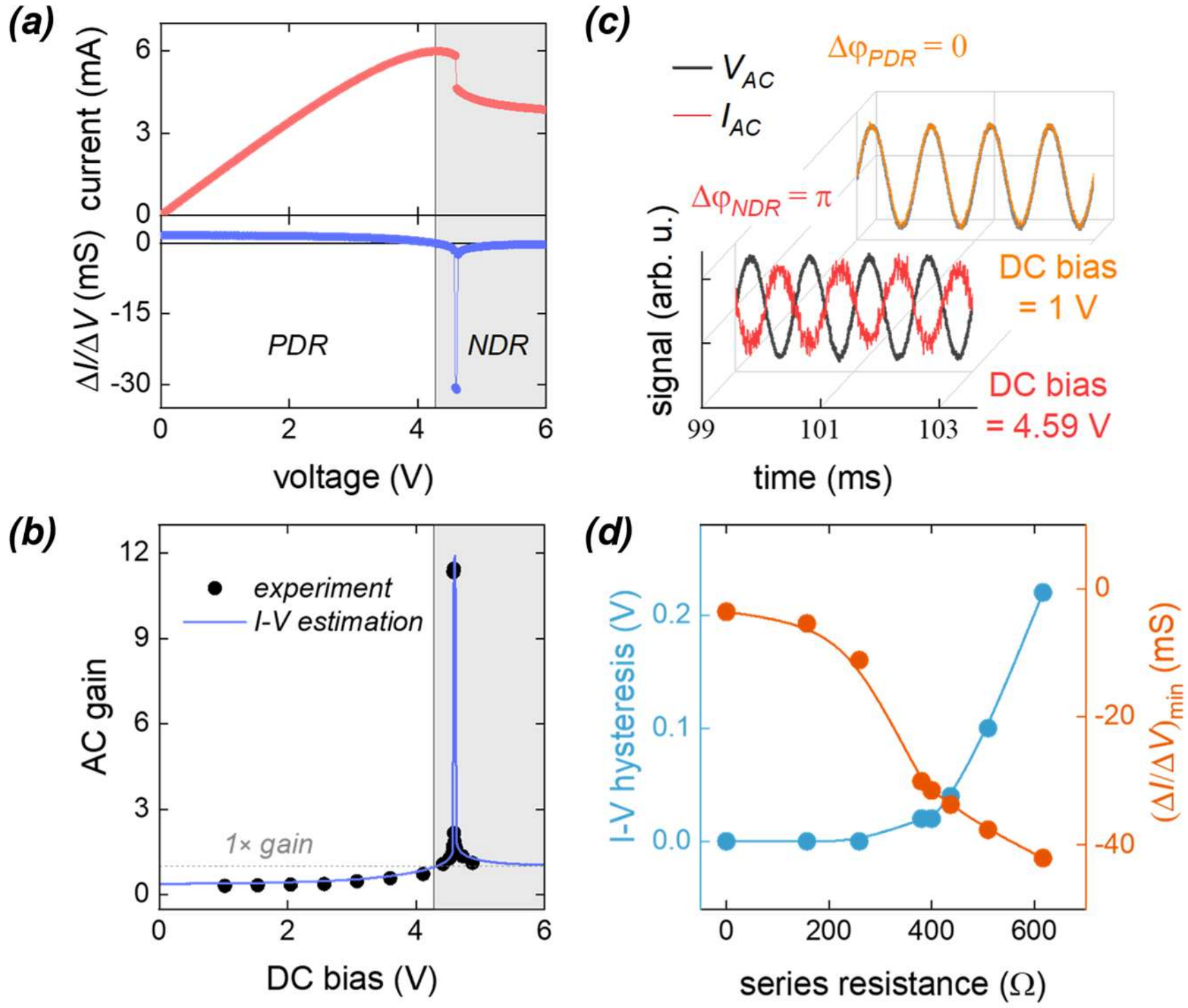

**Fig. 3. a.** The top panel shows the I-V characteristic of LSMO during an increasing voltage ramp. The bottom panel shows the differential conductance. The PDR and NDR regions are highlighted. **b.** Experimental AC gain as a function of DC bias (black symbols) and corresponding gain estimate using Eq. (3) (blue line). **c.** Normalized AC voltage and current recorded in the PDR ($V_{DC}$ = 1 V) and NDR ($V_{DC}$ = 4.59 V) regimes. When the LSMO device is biased in the NDR regime a π phase shift is present between the AC voltage and current. **d.** I-V hysteresis window (blue) and peak differential conductance (orange) as a function of the series resistance, $R_s$. Increasing $R_s$ leads to an increase in $(\Delta I/\Delta V)_{peak}$ but simultaneously increases the switching hysteresis. All measurements were performed at T = 80 K. Data in panels (b–d) were measured using an AC excitation of 40 mV and a series resistance of $R_s$ = 400 Ω was used in panels (a–c).

The above equation matches the experimental data extremely well (**Fig. 3b**, blue line). The above-unity gain is due to the sign reversal of $\Delta I/\Delta V$ when the MIT device is biased close to switching. In the PDR region, where $\Delta I/\Delta V > 0$, the gain is always lower than 1 because the $R_s\cdot(\Delta I/\Delta V)$ term is subtracted in **Eq. (3)**. In the NDR region, where $\Delta I/\Delta V < 0$, the $R_s\cdot(\Delta I/\Delta V)$ flips sign, and the gain becomes greater than 1. The MIT switching in LSMO produces a large and negative $\Delta I/\Delta V$ (**Fig. 3a**, bottom panel), which enables strong amplification. We thus show that, by using the easily measurable $\Delta I/\Delta V$ as an input, we can accurately predict the gain of the MIT amplifier.

Considering the relation between the AC voltage and current provides another perspective on the development of the above-unity gain. **Fig. 3c** shows the measured input AC voltage and AC current when the LSMO device is biased in the PDR ($V_{DC}$ = 1 V) or the NDR ($V_{DC}$ = 4.59 V) regions. All signals are normalized to facilitate comparison. When the device is biased in the PDR region, the input AC voltage and the AC current are perfectly in-phase, resulting in an attenuated output AC voltage (AC gain < 1), as expected in a conventional voltage divider circuit. Because the AC excitation probes the local $\Delta I/\Delta V$ of the I-V characteristic, biasing the device in the NDR region results in a π phase shift between the AC voltage and current. Consequently, an increase in AC voltage leads to a decrease in AC current, which reduces the voltage drop across the series resistor (i.e., $I_{AC}{\cdot}R_s$ decreases). Because the sum of the voltages across the series resistor and the LSMO device must equal the input AC voltage, the decrease in $I_{AC}{\cdot}R_s$ is compensated by an increase in the AC voltage across the LSMO device, which is the output of our amplifier. Thus, the above-unity gain is achieved when the LSMO device is biased in the NDR region, where there is a phase inversion between the AC voltage and current.

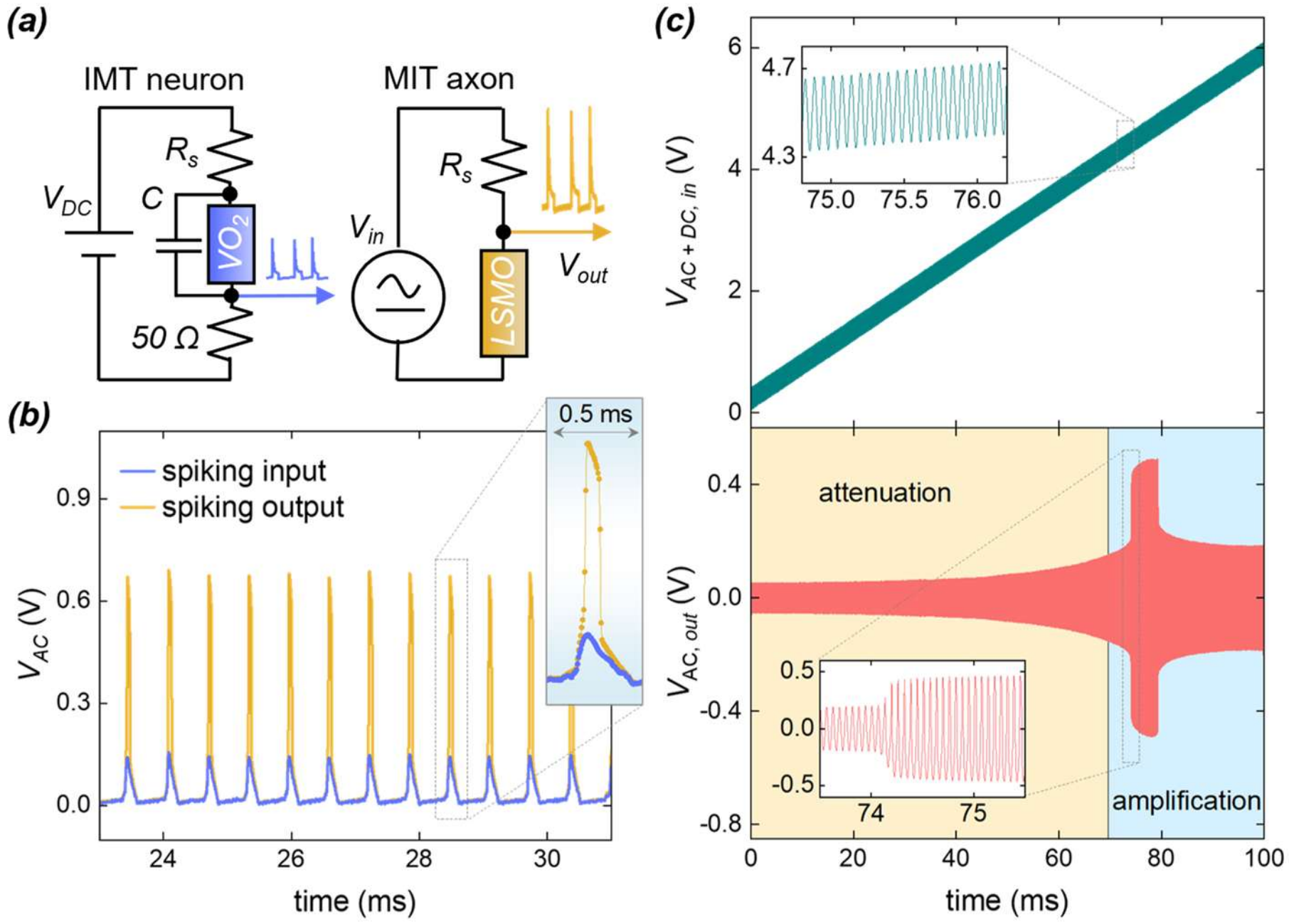

**Fig. 4. a.** Schematic of the experiment in which a neuron-like spiking sequence obtained using a $VO_2$ oscillator is fed into the LSMO amplifier. In this experiment, the LSMO device is DC-biased in the NDR region, and the experimental spiking sequence is superimposed on the DC bias. **b.** Time traces of the AC input (blue) and output (orange) signals, demonstrating the amplification of neuron-like spikes by the MIT axon. **c.** The top panel shows the applied AC perturbation superimposed on a slowly increasing linear voltage ramp. The bottom panel shows the corresponding AC response of the MIT amplifier. The amplifier can be dynamically actuated between the attenuation and amplification regimes by varying the voltage bias. All measurements were performed at T = 80 K using $R_s$ = 400 Ω.

**Eq. (3)** suggests that increasing the series resistance $R_s$ promotes higher gain because the term $R_s \cdot (\Delta I/\Delta V)$ increases and the MIT switching becomes more abrupt (**Fig. 1a**), resulting in a larger $\Delta I/\Delta V$. The results in **Fig. 2b** indeed show a steady amplification enhancement with increasing $R_s$ in the 157 – 400 Ω range. However, this gain enhancement does not continue indefinitely. At larger $R_s$, the I-V characteristic develops switching hysteresis (**Fig. 1a**). **Fig. 3d** shows that while increasing $R_s$ results in a larger $(\Delta I/\Delta V)_{peak}$, it also leads to a wider switching hysteresis. When the I-V characteristic is hysteretic, the amplitude of the input AC excitation must exceed the voltage hysteresis window to drive the device across the switching region during each cycle and achieve a large $\Delta I/\Delta V$. This requirement reduces the achievable gain according to **Eq. (2)**. Overall, amplification optimization requires simultaneous adjustment of several parameters to tune the switching characteristics of the MIT device. From our experiments, the largest gain is achieved when the I-V characteristics exhibit abrupt switching with minimal hysteresis.

As a proof of concept, we demonstrate the amplification of neuron-like electrical spikes. In this experiment (**Fig. 4a**), we first recorded an experimental spiking sequence generated by a $VO_2$-based artificial neuron [22,42] and then fed this sequence into the DC-biased LSMO axon. **Fig. 4b** shows the input and output signals of our amplifier (AC components with the DC bias removed). We achieved an amplification factor of ~5× for the input electrical spikes. Importantly, while there are some minor differences between the input and output signals, the characteristic shape of a neuron-like spike overall is well-preserved by the LSMO amplifier (inset in **Fig. 4b**). Unlike gradually varying sine-wave excitations (**Fig. 1d**), sharp spikes evidently are much less sensitive to signal distortions introduced by the nonlinear electrical properties of the LSMO device. These results, therefore, demonstrate that the MIT-based amplifier is particularly well suited for neuromorphic applications, as it can help prevent information loss in large-scale spiking neural networks by enabling signal amplification locally at a neuron junction.

The gain of the MIT amplifier can be dynamically actuated by adjusting the voltage bias. We demonstrate dynamic gain control by applying an AC perturbation to our LSMO amplifier, superimposed on a slowly increasing linear voltage bias ramp (**Fig. 4c**, top panel). The corresponding AC response of the amplifier is shown in the bottom panel of **Fig. 4c**. Two regimes can be identified in the AC response. First, the attenuation regime (orange-shaded area), which corresponds to the low-voltage region of the slowly increasing bias ramp, during which the LSMO device acts as a conventional resistor. Second, the amplification regime (blue-shaded area), which emerges when the bias ramp reaches a sufficiently high voltage to drive the LSMO device into the onset of the MIT. These results show that varying the bias allows for dynamic control of the transition between signal attenuation and amplification. Such attenuation-amplification control can enable the implementation of basic synaptic functions, such as weighted signal transmission, expanding the range of applications of the MIT amplifier beyond axon emulation.

## Discussion

According to **Eq. (3)**, a necessary condition for achieving an above-unity gain is a negative $\Delta I/\Delta V$, i.e., the presence of an NDR in the I-V characteristic. It is therefore likely that the observed AC signal amplification is not specific to LSMO, and that similar functionalities can be found in a broader class of materials exhibiting strongly nonlinear electrical properties, such as volatile resistive switching. N-type NDR materials, similar to our LSMO, can be directly incorporated into our amplifier circuit. Materials that exhibit an S-type NDR in their I-V characteristics, such as insulator-to-metal transition compounds ($VO_2$, $NbO_2$, $SmNiO_3$, etc.) [22,43–45], could also be used for amplification, provided the circuit is modified to implement current-controlled conditions required for S-type NDR.

We note that nonvolatile memristive materials, such as those exhibiting ionic-migration-based resistive switching, while often have an NDR region in the I-V characteristic [46,47], are not directly suitable for the use in the proposed amplifier circuit. The key property enabling the above-unity gain in our experiments is the ability to dynamically access a portion of the NDR region using a small-amplitude AC excitation. In non-volatile memristors, the NDR region typically appears once during the applied voltage ramp, and the memristor must be reset by applying a large voltage to bring the material into the NRD regime again.

Finding the optimal axon-like amplification material requires a combination of a pronounced dynamically actuated NDR, minimal hysteresis in the I-V characteristic, and stable operation in the NDR regime, i.e., the ability to repeatedly bias the device into a persistent NDR regime for arbitrary time periods. For instance, $VO_2$ is an attractive candidate to achieve amplification, as it exhibits sharp insulator-to-metal switching above room temperature [48]. However, its I-V characteristics often has a pronounced hysteretic behavior [49], which will limit the maximum gain of a $VO_2$-based amplifier. Achieving optimal performance in a $VO_2$-based amplifier necessitates studies to suppress or at least reduce its I-V hysteresis, for example by using focused ion beam irradiation [50].

## Summary

Our results establish a general framework for signal amplification in highly nonlinear electronic phase switching materials. We demonstrate axon-like behavior in an MIT compound, achieving an amplification of ~11.5×. The above-unity gain appears when the device is biased into an NDR regime, which we access by electrical triggering the MIT. Maximizing the NDR, while simultaneously minimizing hysteretic I-V effects, is the key to enhancing amplification. Practically, amplification in our system can be tuned by readily adjustable parameters, including series resistance, AC excitation amplitude, DC bias, and temperature, as these parameters control the electrical switching properties and, consequently, the NDR characteristics of our MIT device. The MIT axon is based on a simple two-terminal device geometry that does not require a large contact area between the active material and electrodes to achieve amplification, which can facilitate scalability. The ability to dynamically switch between attenuation and amplification regimes by changing the DC bias (i.e., operating the circuit either in PDR or NDR regions) allows the MIT device to act as adaptative gain element combining active axonal signal transmission and synaptic weighting functionalities. As a potential practical application, we demonstrate the coupled operation of a $VO_2$-based spiking neuristor and an LSMO-based axon, showing the use of local signal amplification under realistic neuromorphic conditions. Spike amplification may enable the development of large-scale spiking neural networks that are robust against rapid information decay during propagation through multiple layers of dissipative neuronal and synaptic elements.

## Methods

Sample preparation. 20-nm-thick $La_{0.7}Sr_{0.3}MnO_3$ films were epitaxially grown on (001)-oriented $SrTiO_3$ substrates by pulsed laser deposition using a laser fluence of 0.6 J/cm$^2$ and a repetition rate of 1 Hz. The substrates temperature was 700 °C and the oxygen pressure was 300 mTorr during the film growth. After deposition, the films were gradually cooled to room temperature in 300 Torr oxygen pressure. (20 nm Pd)/(20 nm Au) low-contact-resistance electrodes and (20 nm Ti)/(100 nm Au) wire bonding pads were fabricated using standard photolithography and e-beam evaporation. Isolated switching devices were patterned using Ar ion milling.

Electrical measurements. The measurements were performed in a liquid nitrogen cooled Lakeshore Cryotronics TTPX probe station. Quasi-static I-V characteristics were recorded using a Keithley 2450 source meter. Dynamic AC amplification measurements were performed using a Tektronix AFG3011C function generator and a Tektronix MSO54 oscilloscope.

## Acknowledgment

Work at the University of California San Diego and the University of California Davis was supported as part of the Quantum Materials for Energy Efficient Neuromorphic Computing (Q-MEEN-C), an Energy Frontier Research Center funded by the U.S. Department of Energy, Office of Science, Basic Energy Sciences under Award # DE-SC0019273. Work at the University of Denver was supported by the U.S. Department of Energy, Office of Science, Basic Energy Sciences, under Award # DE-SC0026129. This work was performed in part at the San Diego Nanotechnology Infrastructure (SDNI) of UCSD, a member of the National Nanotechnology Coordinated Infrastructure, which is supported by the National Science Foundation (Grant ECCS-2025752).

## References

1. Wang, S. *et al.* Technology Roadmap of Bioinspired Computing Hardware. *ACS Nano* **20**, 8102–8163 (2026).
2. Debanne, D. Information processing in the axon. *Nat. Rev. Neurosci.* **5**, 304–316 (2004).
3. Rama, S., Zbili, M. & Debanne, D. Signal propagation along the axon. *Curr. Opin. Neurobiol.* **51**, 37–44 (2018).
4. Hoffmann, A. *et al.* Quantum materials for energy-efficient neuromorphic computing: Opportunities and challenges. *APL Mater.* **10**, 070904 (2022).
5. Azevedo, F. A. C. *et al.* Equal numbers of neuronal and nonneuronal cells make the human brain an isometrically scaled-up primate brain. *J. Comp. Neurol.* **513**, 532–541 (2009).
6. Kandel, E. R., Koester, J. D., Mack, S. H. & Siegelbaum, S. A. *Principles of Neural Science*. (McGraw-Hill Education, New York, NY, USA, 2013).
7. Ehrenstein, G. & Lecar, H. The Mechanism of Signal Transmission in Nerve Axons. *Annu. Rev. Biophys. Bioeng.* **1**, 347–366 (1972).
8. del Valle, J., Ramírez, J. G., Rozenberg, M. J. & Schuller, I. K. Challenges in materials and devices for resistive-switching-based neuromorphic computing. *J. Appl. Phys.* **124**, (2018).

9. Park, T. J. *et al.* Complex Oxides for Brain-Inspired Computing: A Review. *Adv. Mater.* **35**, 2203352 (2023).
10. Schulman, A., Huhtinen, H. & Paturi, P. Manganite memristive devices: recent progress and emerging opportunities. *J. Phys. Appl. Phys.* **57**, 422001 (2024).
11. Rondinelli, J. M., Caffrey, N. M., Sanvito, S. & Spaldin, N. A. Electronic properties of bulk and thin film $SrRuO_3$: Search for the metal-insulator transition. *Phys. Rev. B* **78**, 155107 (2008).
12. Georgescu, A. B. & Millis, A. J. Quantifying the role of the lattice in metal–insulator phase transitions. *Commun. Phys.* **5**, 135 (2022).
13. Chicca, E., Stefanini, F., Bartolozzi, C. & Indiveri, G. Neuromorphic Electronic Circuits for Building Autonomous Cognitive Systems. *Proc. IEEE* **102**, 1367–1388 (2014).
14. Salahuddin, S., Ni, K. & Datta, S. The era of hyper-scaling in electronics. *Nat. Electron.* **1**, 442–450 (2018).
15. Mehonic, A. & Kenyon, A. J. Brain-inspired computing needs a master plan. *Nature* **604**, 255–260 (2022).
16. Torres, F., Basaran, A. C. & Schuller, I. K. Thermal Management in Neuromorphic Materials, Devices, and Networks. *Adv. Mater.* **35**, 1–23 (2023).
17. Christensen, D. V. *et al.* 2022 roadmap on neuromorphic computing and engineering. *Neuromorphic Comput. Eng.* **2**, 022501 (2022).
18. Schuman, C. D. *et al.* Opportunities for neuromorphic computing algorithms and applications. *Nat. Comput. Sci.* **2**, 10–19 (2022).
19. Indiveri, G. & Liu, S.-C. Memory and Information Processing in Neuromorphic Systems. *Proc. IEEE* **103**, 1379–1397 (2015).
20. Eshraghian, J. K. *et al.* Training Spiking Neural Networks Using Lessons From Deep Learning. *Proc. IEEE* **111**, 1016–1054 (2023).
21. Pickett, M. D., Medeiros-Ribeiro, G. & Williams, R. S. A scalable neuristor built with Mott memristors. *Nat. Mater.* **12**, 114–117 (2013).
22. Yi, W. *et al.* Biological plausibility and stochasticity in scalable $VO_2$ active memristor neurons. *Nat. Commun.* **9**, 4661 (2018).
23. del Valle, J., Salev, P., Kalcheim, Y. & Schuller, I. K. A caloritronics-based Mott neuristor. *Sci. Rep.* **10**, 4292 (2020).
24. Moon, T. *et al.* Leveraging volatile memristors in neuromorphic computing: from materials to system implementation. *Mater. Horiz.* **11**, 4840–4866 (2024).

25. Torrejon, J. *et al.* Neuromorphic computing with nanoscale spintronic oscillators. *Nature* **547**, 428–431 (2017).

26. Grollier, J. *et al.* Neuromorphic spintronics. *Nat. Electron.* **3**, 360–370 (2020).

27. Wang, Z. *et al.* Resistive switching materials for information processing. *Nat. Rev. Mater.* **5**, 173–195 (2020).

28. Jo, S. H. *et al.* Nanoscale Memristor Device as Synapse in Neuromorphic Systems. *Nano Lett.* **10**, 1297–1301 (2010).

29. La Barbera, S., Vuillaume, D. & Alibart, F. Filamentary Switching: Synaptic Plasticity through Device Volatility. *ACS Nano* **9**, 941–949 (2015).

30. Garcia, V. & Bibes, M. Ferroelectric tunnel junctions for information storage and processing. *Nat. Commun.* **5**, 4289 (2014).

31. Lee, S. *et al.* Variance-aware weight quantization of multi-level resistive switching devices based on Pt/$LaAlO_3$/$SrTiO_3$ heterostructures. *Sci. Rep.* **12**, 9068 (2022).

32. Vogel, E. Technology and metrology of new electronic materials and devices. *Nat. Nanotechnol.* **2**, 25–32 (2007).

33. Brown, T. D. *et al.* Axon-like active signal transmission. *Nature* **633**, 804–810 (2024).

34. Brown, T. D., Kumar, S. & Williams, R. S. Physics-based compact modeling of electro-thermal memristors: Negative differential resistance, local activity, and non-local dynamical bifurcations. *Appl. Phys. Rev.* **9**, 011308 (2022).

35. Gao, L., Chen, P.-Y. & Yu, S. $NbO_x$ based oscillation neuron for neuromorphic computing. *Appl. Phys. Lett.* **111**, 103503 (2017).

36. Hennen, T. *et al.* Switching Speed Analysis and Controlled Oscillatory Behavior of a Cr-Doped $V_2O_3$ Threshold Switching Device for Memory Selector and Neuromorphic Computing Application. in *2019 IEEE 11th International Memory Workshop (IMW)* 1–4 (IEEE, Monterey, CA, USA, 2019). doi:10.1109/IMW.2019.8739556.

37. Liu, H. *et al.* A Tantalum Disulfide Charge-Density-Wave Stochastic Artificial Neuron for Emulating Neural Statistical Properties. *Nano Lett.* **21**, 3465–3472 (2021).

38. Adda, C. *et al.* Direct Observation of the Electrically Triggered Insulator-Metal Transition in $V_3O_5$ Far below the Transition Temperature. *Phys. Rev. X* **12**, (2022).

39. Palin, V. *et al.* Inhibitory neuristor based on metal-to-insulator transition. Preprint at https://doi.org/10.48550/arXiv.2604.19951 (2026).

40. Salev, P. *et al.* Transverse barrier formation by electrical triggering of a metal-to-insulator transition. *Nat. Commun.*

**12**, (2021).

41. Salev, P. *et al.* Voltage-controlled magnetic anisotropy enabled by resistive switching. *Phys. Rev. B* **107**, 054415 (2023).
42. Bohaichuk, S. M. *et al.* Fast Spiking of a Mott $VO_2$ –Carbon Nanotube Composite Device. *Nano Lett.* **19**, 6751–6755 (2019).
43. Das, S. K. *et al.* Physical Origin of Negative Differential Resistance in $V_3O_5$ and Its Application as a Solid-State Oscillator. *Adv. Mater.* **35**, (2023).
44. Kumar, S., Strachan, J. P. & Williams, R. S. Chaotic dynamics in nanoscale $NbO_2$ Mott memristors for analogue computing. *Nature* **548**, 318–321 (2017).
45. Luibrand, T. *et al.* Characteristic length scales of the electrically induced insulator-to-metal transition. *Phys. Rev. Res.* **5**, (2023).
46. Du, Y. *et al.* Symmetrical Negative Differential Resistance Behavior of a Resistive Switching Device. *ACS Nano* **6**, 2517–2523 (2012).
47. Feng, Y. *et al.* Negative Differential Resistance Effect in Ru-Based RRAM Device Fabricated by Atomic Layer Deposition. *Nanoscale Res. Lett.* **14**, 86 (2019).
48. Crunteanu, A. *et al.* Voltage- and current-activated metal–insulator transition in $VO_2$ -based electrical switches: a lifetime operation analysis. *Sci. Technol. Adv. Mater.* **11**, 065002 (2010).
49. Murtagh, O., Walls, B. & Shvets, I. V. Controlling the resistive switching hysteresis in $VO_2$ thin films via application of pulsed voltage. *Appl. Phys. Lett.* **117**, 063501 (2020).
50. Ghazikhanian, N. *et al.* Resistive switching localization by selective focused ion beam irradiation. *Appl. Phys. Lett.* **123**, (2023).

# SUPPLEMENTARY INFORMATION

## Signal amplification in simple metal-insulator transition devices

*Victor Palin[1], Nareg Ghazikhanian[1], Matthew Frame[2], Yayoi Takamura[2], Ivan K. Schuller[1], Pavel Salev[3]*

[1]*Department of Physics, University of California San Diego*
[2]*Department of Materials Science & Engineering, University of California Davis*
[3]*Department of Physics & Astronomy, University of Denver*

### A. Origin of AC amplification and gain derivation

The origin of the AC amplification can be understood considering the voltage divider in which the input voltage is distributed between the series resistor $R_s$ and the LSMO device. Under DC bias and AC excitation, the voltage distribution is

$$V_{DC,\,in} + \Delta V_{AC,\,in} = \left(V_{DC,\,R_s} + \Delta V_{AC,\,R_s}\right) + \left(V_{DC,\,LSMO} + \Delta V_{AC,\,LSMO}\right)$$

where $V_{DC}$ is the DC component of the signal and $\Delta V_{AC}$ corresponds to the small AC perturbation. Because the DC signal components are time independent by definition, the above equation also implies the following relation between the AC components

$$\Delta V_{AC,\,in} = \Delta V_{AC,\,R_s} + \Delta V_{AC,\,LSMO}$$

which is equivalent to removing DC offsets in our experiments (see **Suppl. Fig. S2** and main text **Fig. 1d**). Because $R_s$ is a constant series resistor, the AC component of the voltage across the resistor can be replaced by

$$\Delta V_{AC,\,R_s} = \Delta I_{AC} \cdot R_s$$

which leads to

$$\Delta V_{AC,\,in} = \Delta I_{AC} \cdot R_s + \Delta V_{AC,\,LSMO}$$

By defining AC gain as the ratio $\left(\Delta V_{AC,\,LSMO}/\Delta V_{AC,\,in}\right)$, the above equation gives

$$AC\ gain = \frac{\Delta V_{AC,\,LSMO}}{\Delta V_{AC,\,in}} = 1 - \left(\frac{\Delta I_{AC}}{\Delta V_{AC,\,in}}\right) \cdot R_s$$

The above-unity gain, therefore, can be obtained if $\left(\Delta I_{AC}/\Delta V_{AC,\,in}\right) < 0$, i.e., when the LSMO device is biased into the negative differential resistance region.

## B. Additional Figures and discussions

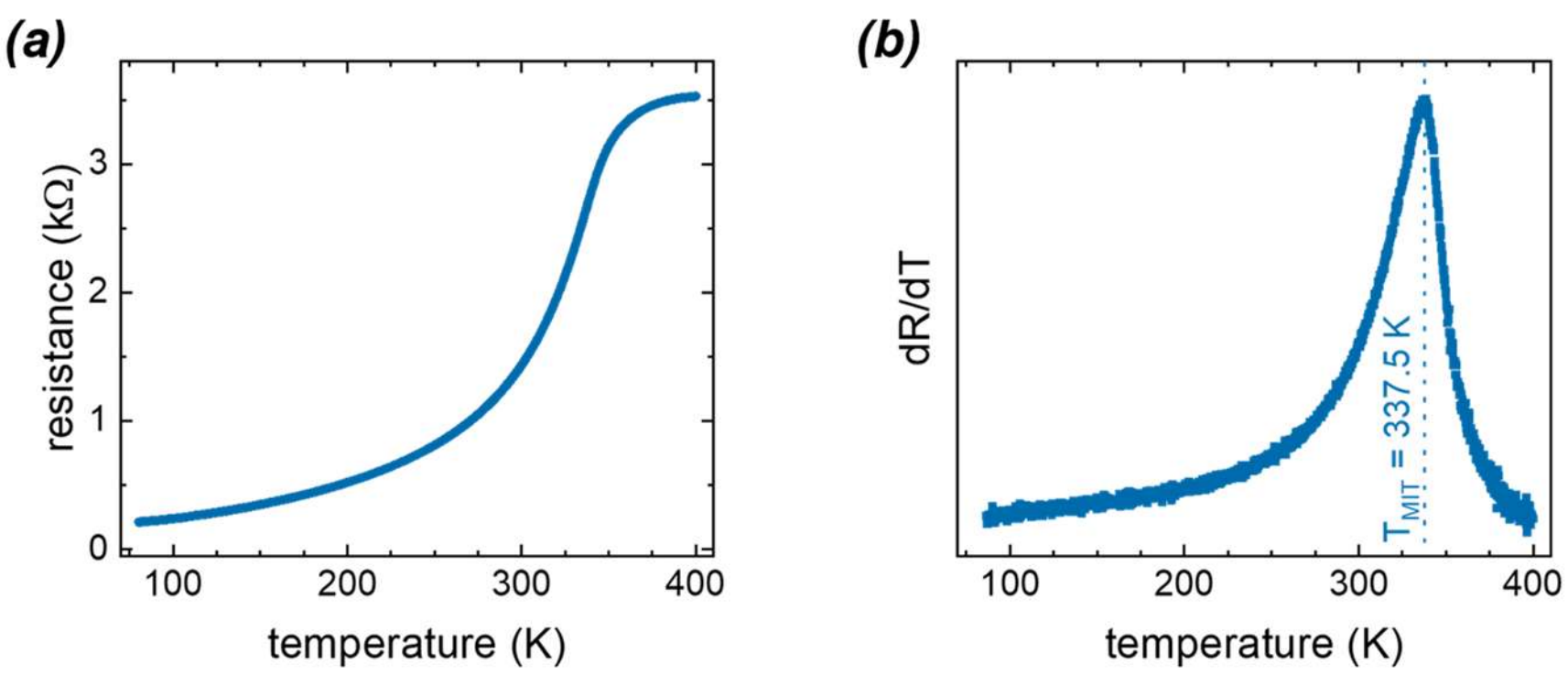

**Supplementary Fig. S1. a.** Typical resistance-temperature dependence of an LSMO device exhibiting the MIT. **b.** Resistance derivative with respect to temperature in which the peak corresponds to the transition temperature of 337.5 K. The measurement was performed using a 2 x 1 µm² device.

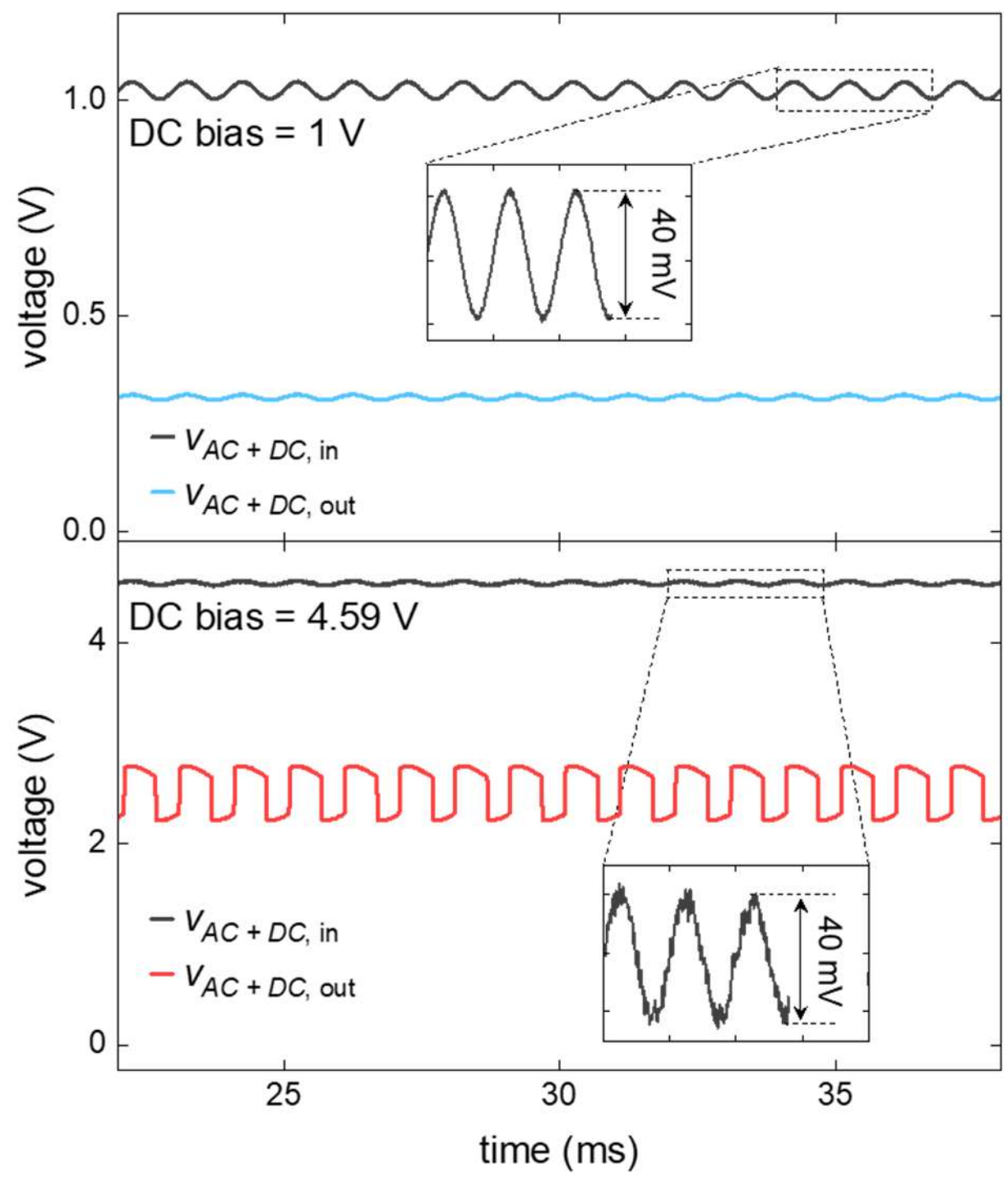

**Supplementary Fig. S2.** Raw input and output signals recorded using a DC bias of 1 V (top panel) and 4.59 V (bottom panel) superimposed on a 40 mV AC excitation. When the LSMO device is biased far away from the MIT switching ($V_{DC}$ = 1 V), the AC output signal is attenuated. When the LSMO device is biased at the MIT switching threshold ($V_{DC}$ = 4.59 V), the AC output signal is strongly amplified. Insets show that both measurements were performed using the same 40 mV AC excitation amplitude. The amplification measurements were performed at 80 K using $R_s$ = 400 Ω on a 2 x 1 µm² device.

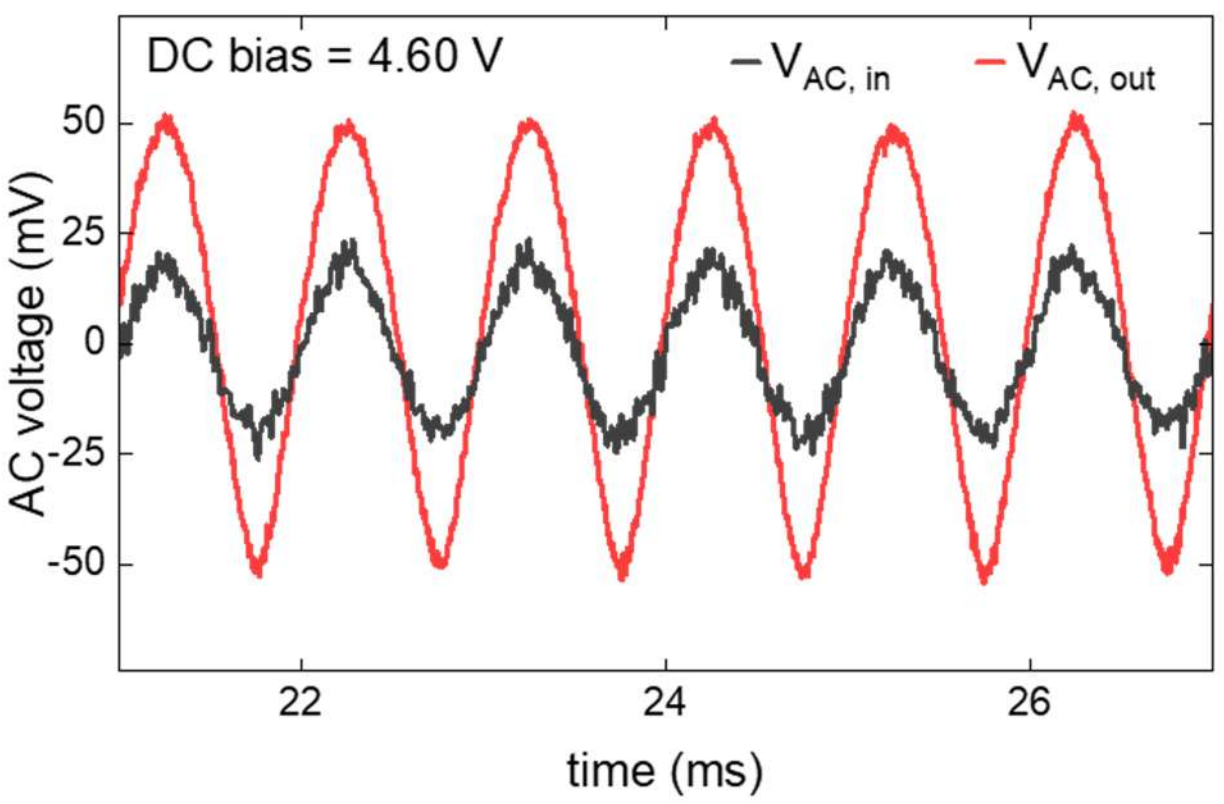

**Supplementary Fig. S3.** Input (black) and output (red) AC signals recorded using a DC bias of 4.60 V and an AC excitation of 40 mV. In this regime, the AC signal is still amplified (~ 2.2×), while distortions are minimized because the LSMO is biased deep in the NDR region. The amplification measurements were performed at 80 K using $R_s$ = 400 Ω on a 2 x 1 μm² device.

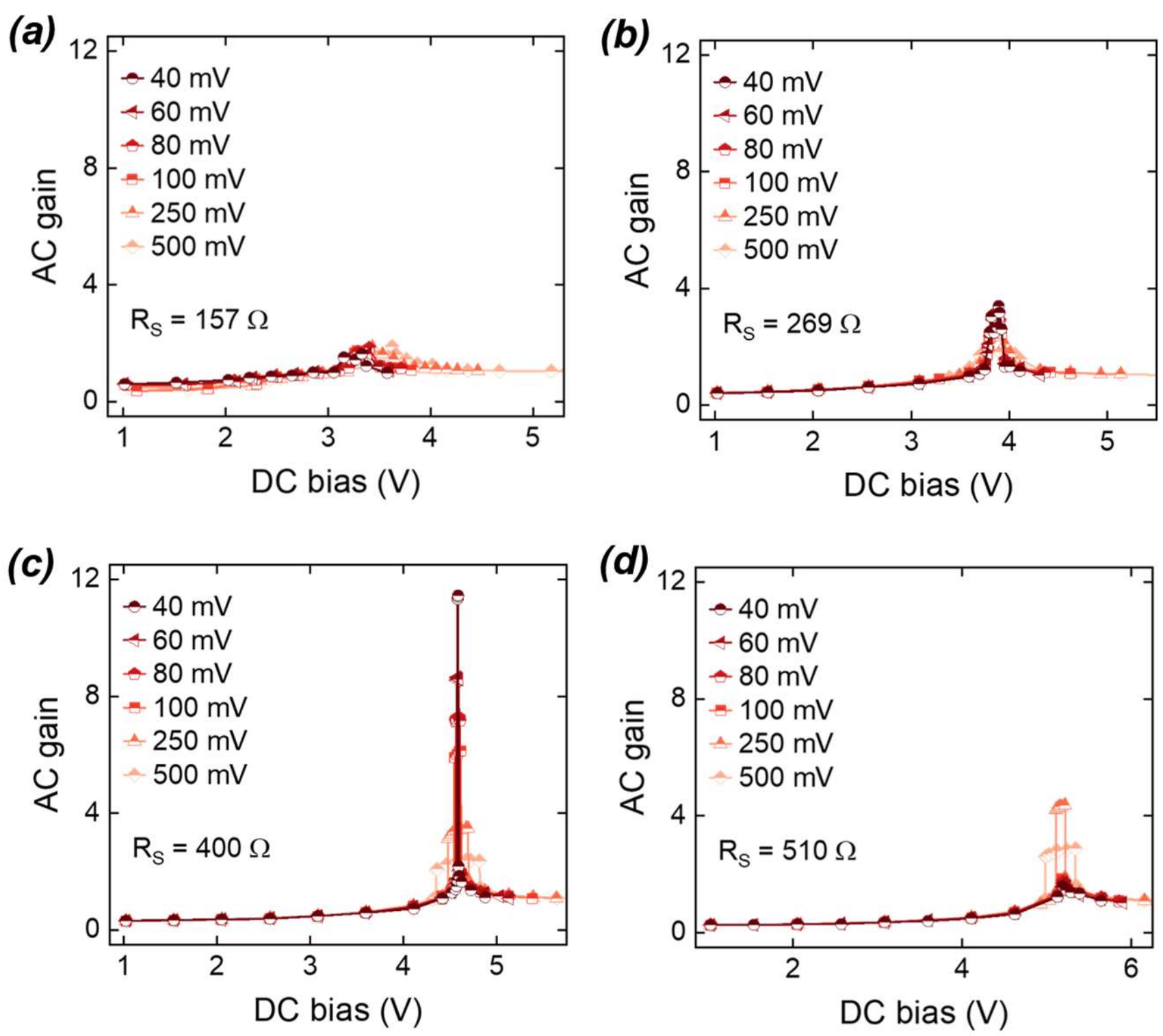

**Supplementary Fig. S4.** AC gain as a function of DC bias for different AC excitation amplitude and for $R_S$ values of 157, 269, 400, and 510 Ω, respectively, in panels (a), (b), (c), and (d). All measurements were performed at T = 80 K using a 2 x 1 μm² device.

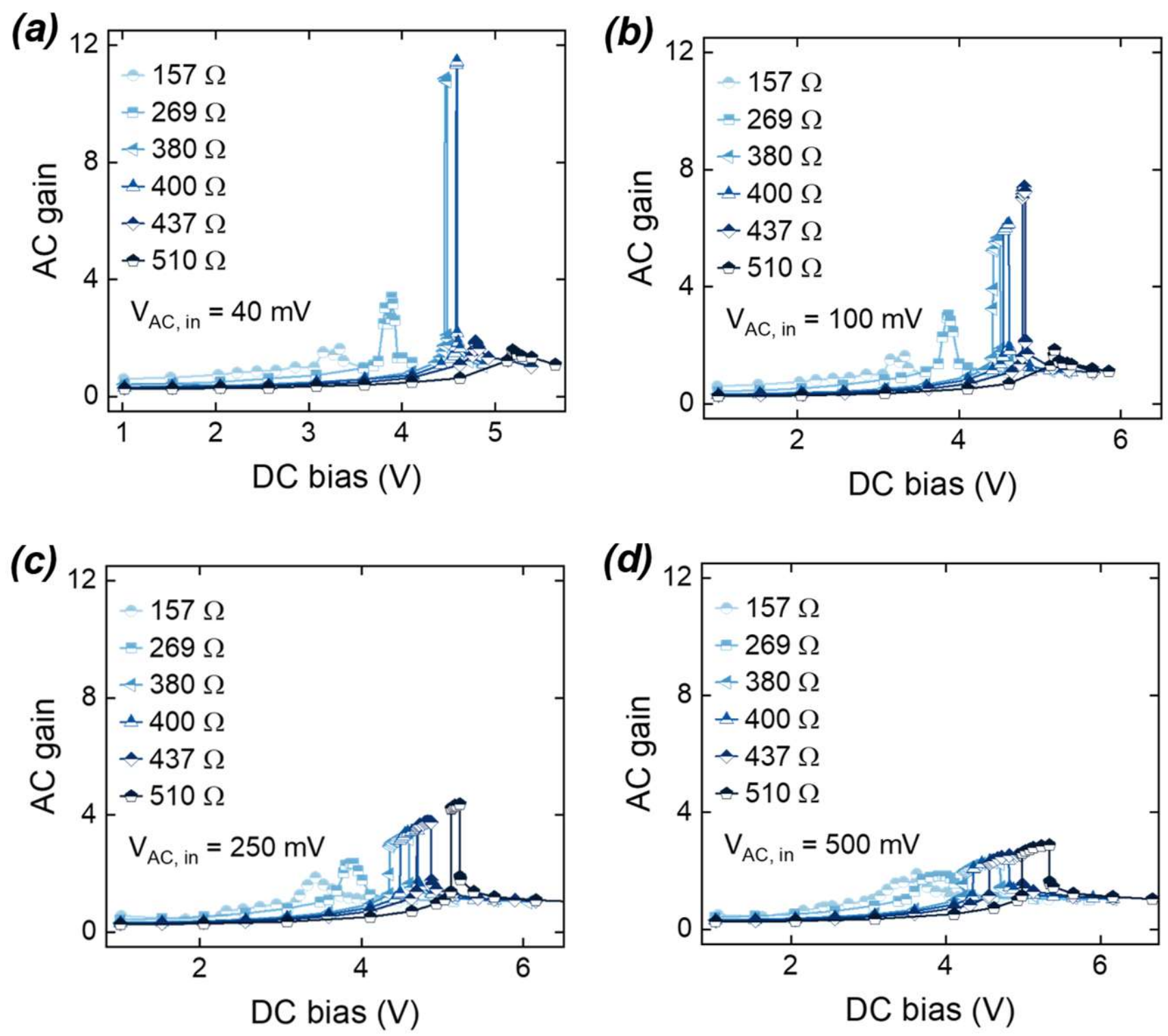

**Supplementary Fig. S5.** AC gain as a function of DC bias for different series resistors and for $V_{AC,in}$ values of 40, 100, 250, and 500 mV, respectively, in panels (a), (b), (c), and (d). All measurements were performed at T = 80 K using a 2 x 1 μm² device.